\documentclass{pasj00}
\draft

\begin{document}
\SetRunningHead{Okuda \& Iguchi}{Performance Measurements of 8-Gsps 1-bit ADCs.}
\Received{2007/08/21}
\Accepted{2007/10/23}

\title{Performance Measurements of 8-Gsps 1-bit ADCs 
Developed for Wideband Radio Astronomical Observations}

 \author{%
   Takeshi \textsc{Okuda}\altaffilmark{1}
   and  
   Satoru \textsc{Iguchi}\altaffilmark{1}
 }
 \altaffiltext{1}{National Astronomical Observatory of Japan, 2-21-1 Osawa, 
 Mitaka, Tokyo 181-8588, Japan}
 \email{takeshi.okuda@nao.ac.jp}
 \email{s.iguchi@nao.ac.jp}


\maketitle

\begin{abstract}
8-Gsps 1-bit Analog-to-Digital Converters (ADCs) were newly developed toward the 
realization of the wideband observation. 
The development of the wideband ADCs is one of the most essential developments 
for the radio interferometer. 
To evaluate its performance in interferometric observations, the time (phase) 
stability and frequency response were measured with a noise source and a signal 
generator. 
The results of these measurements show that the developed ADCs can achieve the 
jitter time less than 0.05 psec at the time interval of 1 sec, the passband 
frequency response with the slope less than 0.73 dB/GHz and the ripple less 
than 1.8 dB, and the aperture time less than 20 psec. 
The details of the developed ADC design, the measurement methods, and the 
results of these measurements are presented in this paper.
\end{abstract}

\section{Introduction}

A high-speed Analog-to-Digital Converter (ADC) can greatly contribute to the 
frontier of radio astronomy. 
The performance of ADC is very important because it decides the sensitivity of 
radio astronomical interferometric observations and the accuracy of astrometry 
observations and geodetic experiments. 
To realize the wideband observation for the Atacama Large 
Millimeter/Submillimeter Array (ALMA), the 2-bit ADC operating at the 4-GHz 
sampling rate have been developed \citep{oki02}. A fringe spectrum that contains 
20 line features as well as continuum emission at $\sim$86.2 GHz was obtained 
though an interferometric observation of the Orion-KL regions using these ADCs 
connected to the Nobeyama Millimeter Array (NMA). 
This result shows that the realization of high frequency-resolution over a wide 
bandwidth is technically feasible \citep{oku02}. 
This is the first time to obtain such a wideband astronomical fringe from a 
\emph{single} correlation product of digitized signals; 20 line features can be 
found by a careful inspection of the spectrum.

In this paper, the development of newly-developed 8-Gsps 1-bit ADC is described 
in Section 2, the measurement methods and the performance of the ADC in 
Section 3, and the conclusion in Section 4. 

\section{The development of 8-Gsps 1-bit ADC}
\begin{figure}
 \begin{center}
   \FigureFile(80mm,80mm){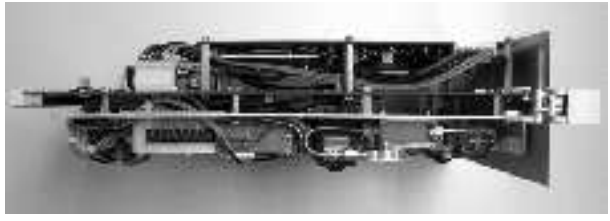}
 \end{center}
 \caption{Top view of the 8-Gsps 1-bit ADC board. The top board, middle board, 
 and bottom board are a bias circuit board, a digital processing board, and an 
 ADC extended board, respectively.}
 \label{fig:01}
\end{figure}

\begin{figure}
 \begin{center}
   \FigureFile(80mm,80mm){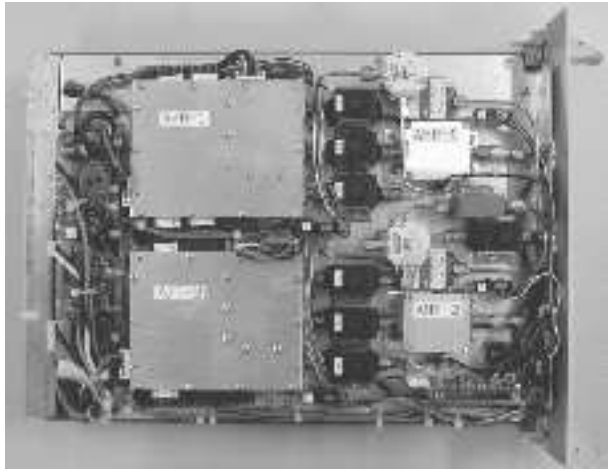}
 \end{center}
 \caption{Side view of the 8-Gsps 1-bit ADC board. 
 This is top view of the ADC extended board (see Fig.~\ref{fig:01}. 
 ``A/D-1'' and ``A/D-2'' are the ADC units that have a built-in DEC and a DMUX.}
 \label{fig:02}
\end{figure}

\begin{figure}
 \begin{center}
   \FigureFile(160mm,160mm){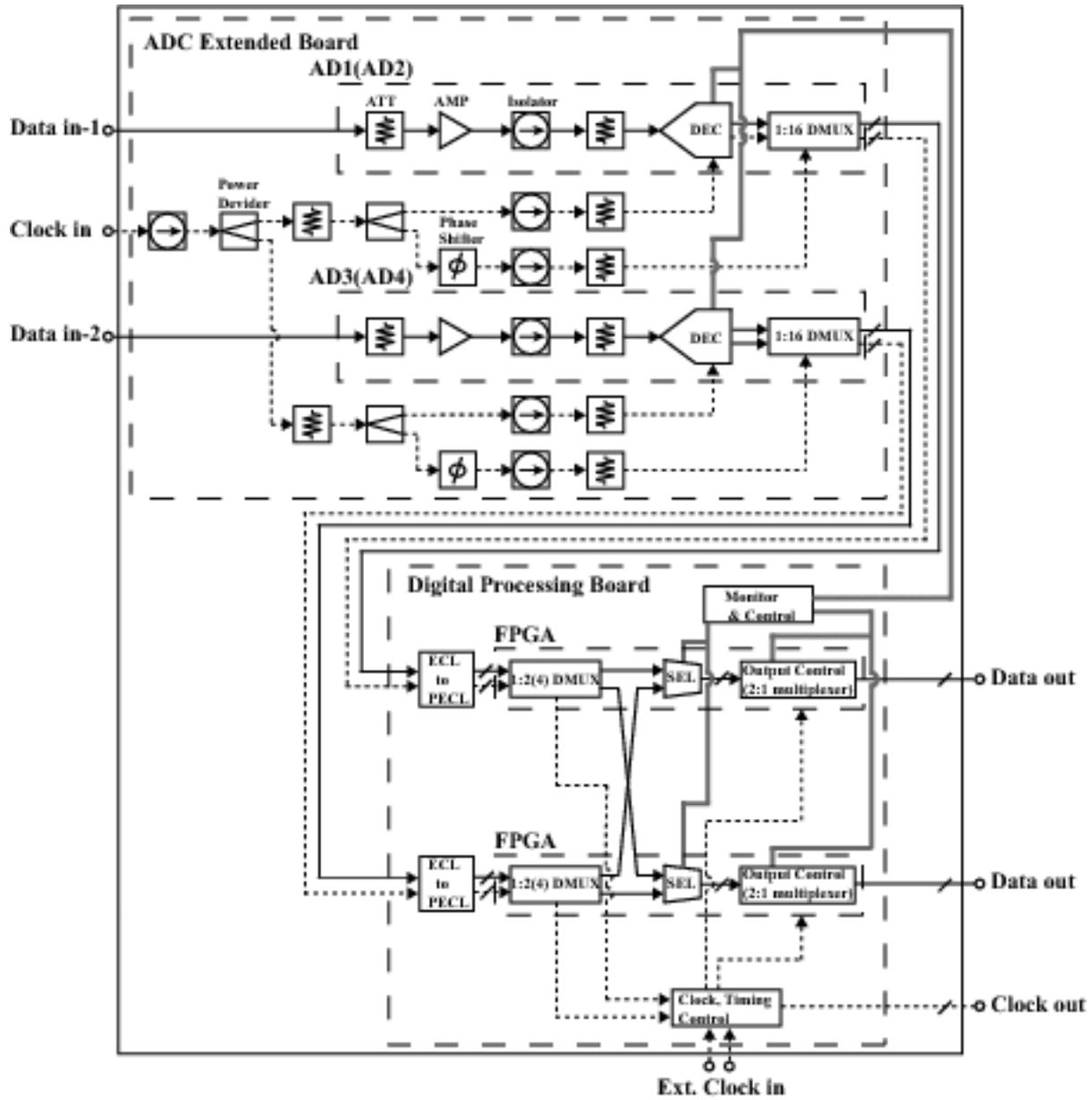}
 \end{center}
 \caption{Functional system block diagram of the 8-Gsps 1-bit ADC board. 
 The solid, dashed, and gray lines indicate IF signals, clock signals, and 
 ``Monitor \& Control" buses, respectively. }
 \label{fig:03}
\end{figure}

\begin{figure}
 \begin{center}
   \FigureFile(80mm,80mm){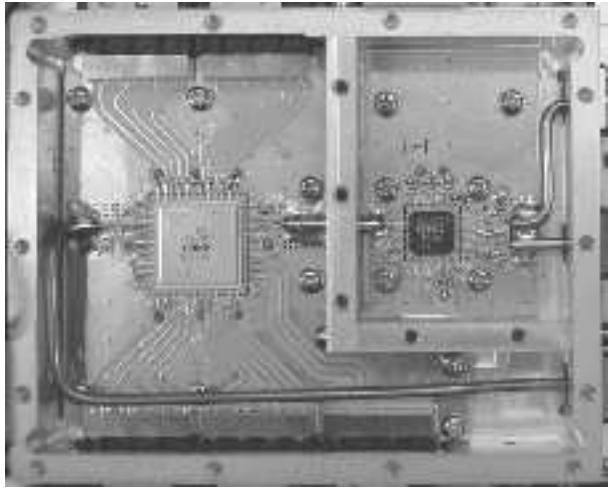}
 \end{center}
 \caption{Inner photograph of the 8-Gsps 1-bit ADC unit that is in the ADC 
 extended board (see Fig.~\ref{fig:02}). 
 The left and right chips are a DMUX and DEC, respectively.}
 \label{fig:04}
\end{figure}

The 8-Gbps 1-bit ADC was developed with commercially available DEcision Circuit 
(DEC) and 1:16 DeMUltipleXer (DMUX) which are currently used as optical 
communication GaAs IC chips at a bit rate of 10 Gbps. 
These chips, DEC (OKI KGL4215) and DMUX (OKI K422C), have been developed for 
high-speed optical communication systems. 
Fig.~\ref{fig:01} and Fig.~\ref{fig:02} shows the developed ADC board that has 
two 8-Gsps 1-bit ADC units. 
Fig.~\ref{fig:03} shows the functional system block diagram of the ADC board. 
Fig.~\ref{fig:04} shows the inner structure of the ADC unit. 

\begin{figure}
 \begin{center}
   \FigureFile(143mm,160mm){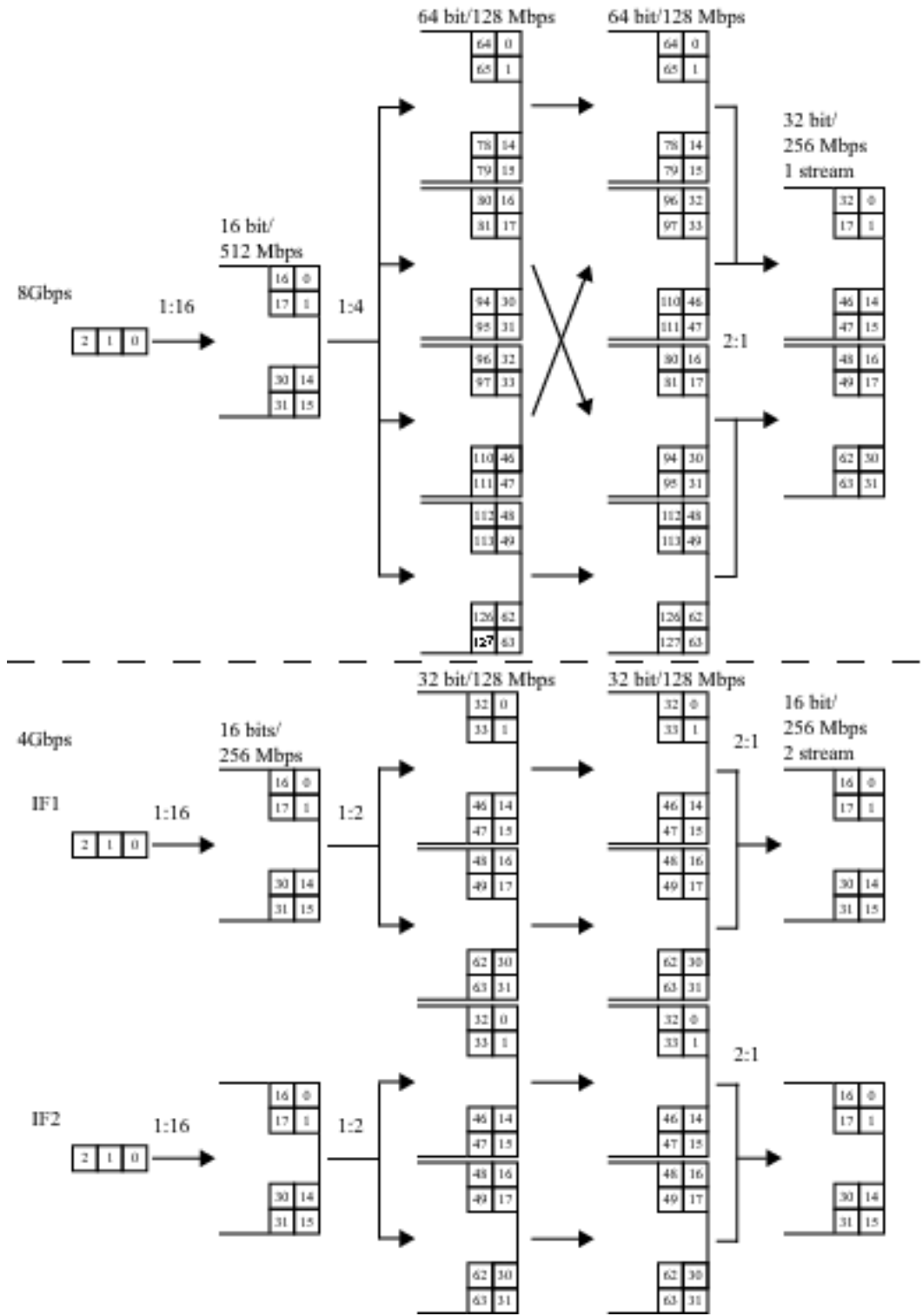}
 \end{center}
 \caption{Data formats of the developed ADC board according to 
 sampling rates. } 
 \label{fig:05}
\end{figure}

The input signal power level is -20 dBm at the ADC board end. 
In this ADC board, the input analog signals with a power of -20 dBm are 
amplified with AMP (Quinstar Technology QLJ-04082530-JO), and then the power of 
the input signals to ADC units becomes about $0$ dBm. After being amplified, 
the analog signals are sample-held and quantized at a sampling rate of 8192/4096 
MHz with GaAs DEC chips, and then the quantized waveforms are digitized and 
demultiplexed with 1:16 DMUX chips in the ADC units. 
Since the GaAs DEC chips have a response speed of 10 Gsps, the newly-developed 
ADC will be capable of covering the frequency range from 0 to 10 GHz. 
From the ADC units, the digital signals paralleled to 16 bits by 1:16 DMUX chips 
are output to the digital processing board (Fig. \ref{fig:04}). 
On the digital processing board, the digital signal level is first converted 
from ECL (Emitter Coupled Logic) to PECL (Positive Referenced Emitter Coupled 
Logic). 
In a FPGA (Field Programmable Gate Array), the data stream is adjusted and 
unified to be 64 (32 $\times$ 2) bits $\times$ 128 Mbps at both of the 4-and 
8-GHz sampling rates by switching the 1:2 and 1:4 demultiplexers, and by 
selecting the appropriate signal paths (see Fig.~\ref{fig:05}). 
After being selected, the digital signals are multiplexed by 2:1 multiplexer in 
``Output Control" in FPGA as shown in Fig.~\ref{fig:03}, and then the output 
data stream from the ADC board is 32 bits $\times$ 256 Mbps as shown in 
Fig.~\ref{fig:05}. 
The relationship between the sampling rate, the decimation factor and output 
data lines is listed in Table \ref{tab:sampling}.

A reference clock signal (input level of $+10$ dBm) at 4 or 8 GHz is divided 
into two ADC units, and then the clocks are provided to the DEC and DMUX chips 
(see Fig. \ref{fig:03}). 
In this case, there is a delay caused between the DEC and DMUX chips because the 
transmission path length is different (see Fig.~\ref{fig:04}). 
The delay is adjusted by the phase shifter (Anritsu Corp. A5N1102) in the ADC 
board. 
The ADC board also outputs the reference clocks of IF1 and IF2 which are 
synchronized with digital signals. 
Although these clocks are generated from the output signals of 1:16 DMUX chip, 
the phase of IF1 is not coherent with that of IF2 because they are transmitted 
via different communication paths. 
In this ADC board, this phase skew is absorbed in ``Output Control'' by selecting 
either of the reference clocks in ``Clock, Timing Control" (Fig.~\ref{fig:03}). 
Also, this internal reference clock can be generated and synchronized with an 
external clock by selecting it in ``Clock, Timing Control".

Bit distributions in the Analog-to-digital conversion can be estimated on the 
FPGAs. 
In the ADC board, the bit distributions are obtained on each of the 16 parallel 
lines (bits) or on the sum of all bits. 
The results of the bit distribution measurement can be read by accessing an 
internal Monitor $\&$ Control unit which is integrated in the ADC board. 
Also, to adjust the threshold level, the analog offset of the ADC board can be 
changed by a command issued from the Monitor $\&$ Control computer to the DEC 
chips via the Monitor $\&$ Control unit (Fig.~\ref{fig:03}).

\section{Measurement of ADC Performance}
Phase stability of the ADC was evaluated from derivation of the Allan variance 
by measuring a phase variation between the stable reference signals and the 
digital-to-analog converted signals \citep{mat95}. 
However, since the measured results included the effects of the time skew in 
digital signal processing and the stability of the DAC (Digital-to-Analog 
Converter), it was difficult to directly evaluate the phase stability or jitter 
time of the ADC.  

The measurement methods and results of the jitter time, the passband frequency 
response and the aperture time of thedeveloped ADCs are presented in this section. 

\subsection{ADC Measurement System}

\begin{figure}
 \begin{center}
  \FigureFile(160mm,160mm){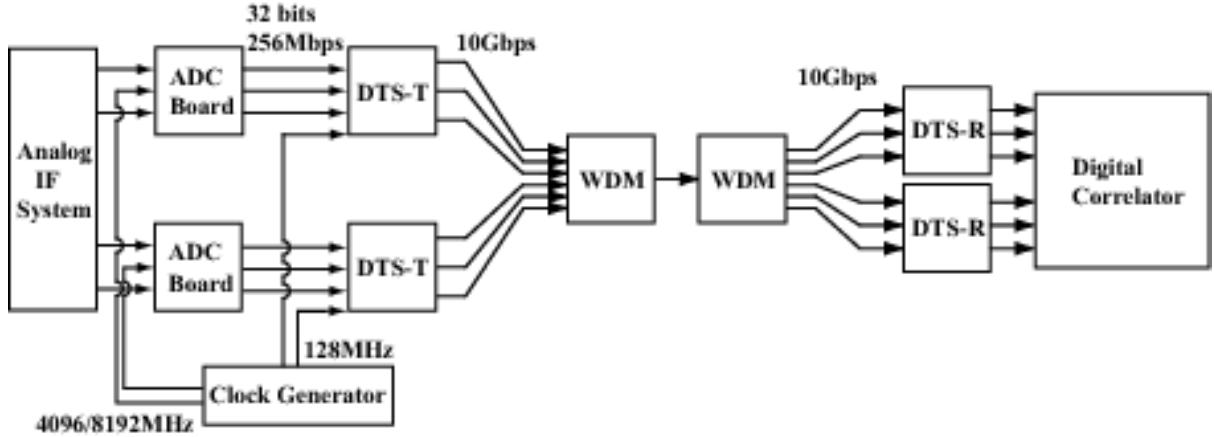}
 \end{center}
 \caption{The system block diagram of the ASTE wideband digital backend system.} 
 \label{fig:06}
\end{figure}

Digital backend system of the wideband IF system has been developed for the 
Atacama Submillimeter Telescope Experiment (ASTE) that is a project to install 
and operate a 10-m submillimeter telescope at a high altitude site (4,800 m) in 
the Atacama desert, Northern Chile \citep{eza04}. 
This system is composed of a reference clock generator, high-speed ADCs, a 
60-Gbps Data Transmission System (DTS), and a digital correlator that can 
operate as an 8-Gsps 3-bits correlator between four input signals. 
The system block diagram of the ASTE wideband digital backend system is shown in 
Fig.~\ref{fig:06}. 

Two types of reference clocks (8192/4096 MHz and 128 MHz), which are provided by 
a high-stability reference clock generator, lock the phase between two ADC units. 
DTS Transmitter (DTS-T) module including the developed ADC boards converts the 
astronomical analog signals into the optical transmission signals of 10.24 Gsps 
$\times$ 3 lines $\times$ 2 IFs. 
In this digital backend system, the WDM (Wavelength-Division Multiplexing) 
modules with 16 wavelengths is used for transmission of these optical signals. 
Of 16 wavelengths, only six wavelengths are used for the DTS system. 
As shown in Fig. 6, the six optical signals are output from the DTS-T modules 
through the WDM modules to the DTS Receiver (DTS-R) boards. 
The WDM modules are connected to each other with only one optical fiber cable. 
The WDM module on the digital correlator side is directly connected to the DTS-R 
boards with 6 optical fiber cables.

\subsection{Allan Variance Measurement}

The fraction frequency fluctuation averaged over the time interval $\tau$ is 
\begin{equation}
\bar{y}_k = \frac{\phi(t_k+\tau)-\phi(t_k)}{2\pi \nu_0 \tau}, 
\end{equation}
where $\phi(t)$ is a phase difference between the two ADCs, $\nu_0$ is a frequency 
of an oscillator. 
The Allan variance is defined as 
\begin{eqnarray}
\sigma_y^2(\tau)
&=& \frac{\left< \left( \bar{y}_{k+1} - \bar{y}_k \right)^2 \right>}{2} \nonumber \\ 
&=& \frac{\left< \left[ \phi(t+2\tau)-2\phi(t+\tau)+\phi(t)  \right]^2\right>}{8 \pi^2 \nu_0^2 \tau^2}.  
\end{eqnarray}

\begin{figure}
 \begin{center}
  \FigureFile(80mm,80mm){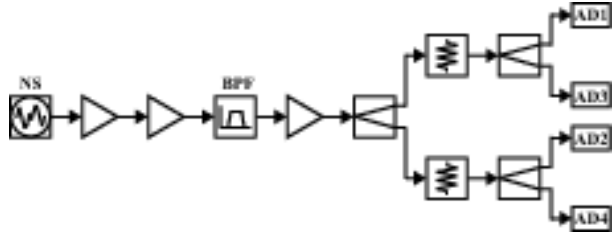}
 \end{center}
 \caption{Block diagram of the instrument setup for Allan variance measurements. 
 NS is the noise source (Agilent Technologies 346B) and BPF is the 4-8GHz 
 bandpass filter.}
 \label{fig:07}
\end{figure}

\begin{figure}
 \begin{center}
  \FigureFile(160mm,160mm){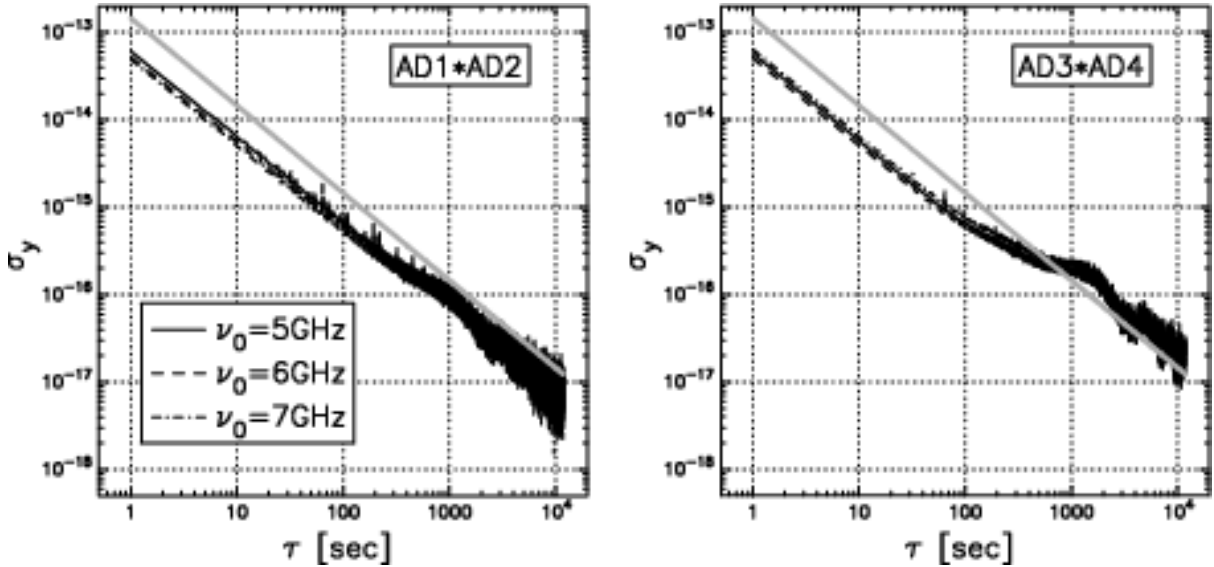}
 \end{center}
 \caption{Allan standard deviation between samples on the second time scale. 
 Figures show Allan standard deviations of AD1$\ast$AD2 and those of AD3$\ast$AD4. 
 Solid lines, dashed lines, and dot lines indicate the results of an IF frequency 
 of 5 GHz, 6 GHz, and 7 GHz, respectively. The gray lines indicate the coherence 
 $\left< C^2(T) \right>$ of 0.99999 in the case of white-phase noise.}
 \label{fig:08}
\end{figure}

The instrument setup for measuring the Allan variance between the two 
developed ADCs is shown in Fig.~\ref{fig:07}. 
The noise signals generated from a noise source (Agilent Technologies 346B) are 
cut with a analog bandpass filter of 4 to 8 GHz. 
The limited-bandwidth noise signals are input to two ADCs. 
The phase difference between two ADC units is measured with the digital correlator. 
The obtained auto-correlations and cross-correlations were corrected from the 
Van Vleck relationship for 1-bit (two-level) quantization \citep{van66} as follows, 
\begin{equation}
\rho^\ast = \frac{2}{\pi} \sin^{-1} \rho, 
\label{eq:van-vleck}
\end{equation}
where $\rho$ is real and $\rho^\ast$ are measured correlation coefficients,
respectively. 
The measurement results of the Allan standard deviation are shown in 
Fig.~\ref{fig:08}.

\begin{figure}
 \begin{center}
  \FigureFile(160mm,160mm){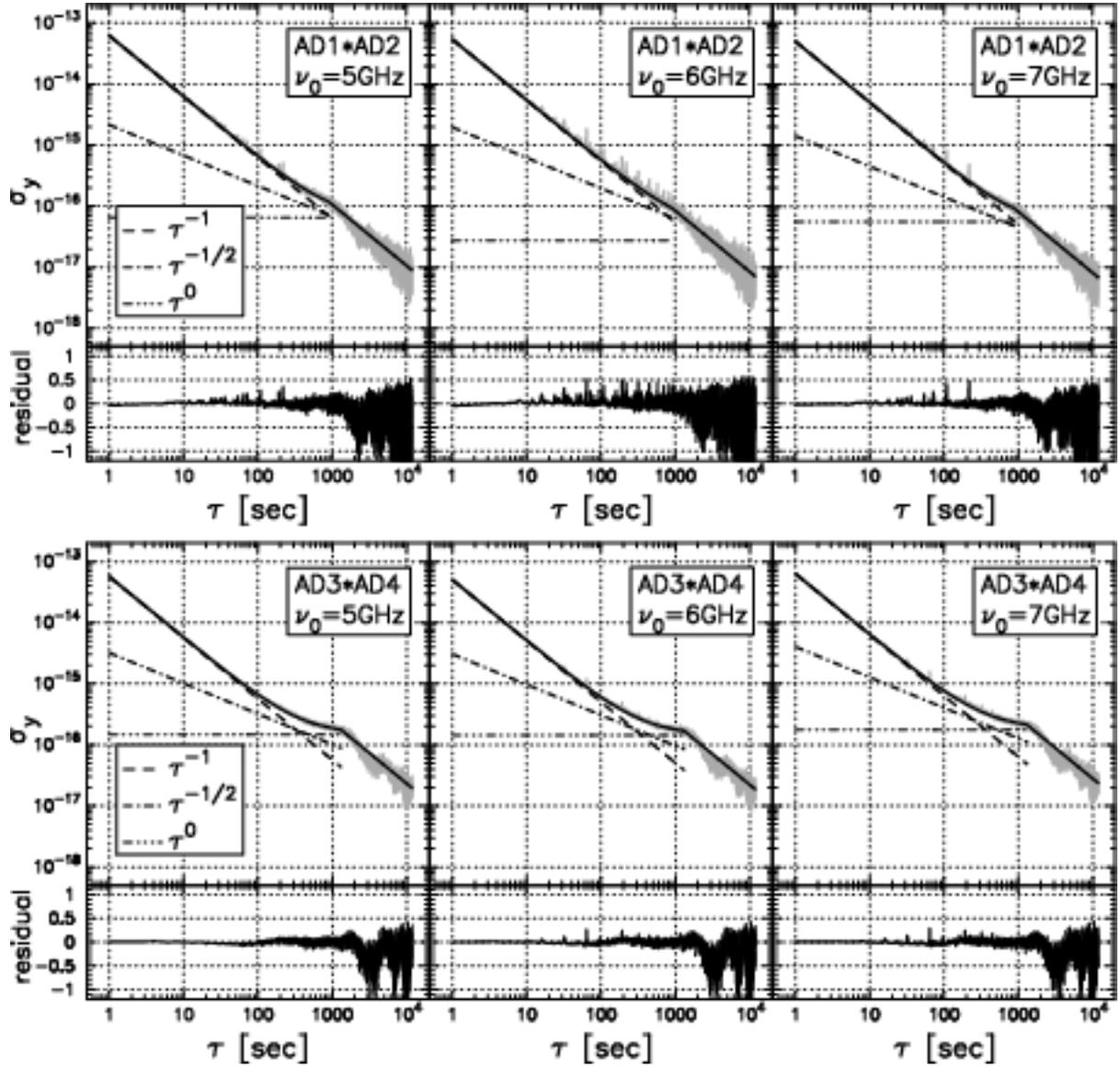}
 \end{center}
 \caption{Results of  least squares fitting to the derived Allan standard 
 deviations. 
 The top panels show the Allan standard deviations (gray) and fitting curves. 
 The solid lines indicate the sum total of the components, $\tau^{-1}$(dashed), 
 $\tau^{-1/2}$(dot-dashed), and $\tau^0$(3dots-dashed). 
 The bottom panels show values of relative residual 
 ($=(\sigma_y-\sigma^\ast)/\sigma_y$, where $\sigma^\ast$ is a derived fitting 
 curve).}
 \label{fig:09}
\end{figure}

The Allan variance can be written as 
\begin{equation} 
\sigma_y^2(\tau) =  \sum_{i=0}^{N_{A}} A_{i}^2 \tau^{-2} 
+ \sum_{i=0}^{N_{B}}B_{i}^2 \tau^{-1} 
+ \sum_{i=0}^{N_C}C_i^2 + \sum_{i=0}^{N_D}D_i^2 \tau, 
\label{eq:allan}
\end{equation} 
where $A_i$, $B_i$, $C_i$ and $D_i$ are the Allan variance in 1 second. 
$A_i$ is due to the white-phase noise and the flicker phase noise, 
$B_i$ is due to the random-walk phase noise and the white-frequency noise, 
$C_i$ is due to the flicker frequency noise, 
and $D_i$ is due to the random-walk frequency noise. 
The slope of measured Allan variance (see Fig.~\ref{fig:08}) indicates that the 
time stability of the developed ADC is dominated by the white-phase noise 
throughout the entire time scale. 
Then Allan variance can be considered as 
\begin{equation}
\sigma_y^2(\tau) =  A_0^2 \tau^{-2} + B_0^2 \tau^{-1} + C_0^2,   
\label{eq:allan-m}
\end{equation}
and these coefficients are derived from a least square fit of these curves. 
The time range for the fit is from 1 sec to $9.0\times10^2$ sec and from 
$9.0\times10^2$ sec to $1.5\times 10^3$ sec for AD1$\ast$AD2 and from 1 sec to 
$1.3\times10^3$ sec and from $1.3\times10^3$ sec to $2.3\times 10^3$ sec for 
AD3$\ast$AD4, but the fit is not performed in more than $\sim 1.5\times 10^3$ sec for 
AD1$\ast$AD2 and more than $\sim 2.3\times 10^3$ sec for AD3$\ast$AD4, 
because the curves of the derived Allan standard deviation turns a steep slope. 
The results of the least squares estimation are shown in Fig.~\ref{fig:09}, 
and then these coefficients are derived as listed in Table \ref{tab:allan}. 

\citet{rogers} defined the coherence function as 
\begin{equation}
C(T) = \left| \frac{1}{T}\int^T_0 e^{j\phi(t)}dt \right|, 
\end{equation}
where $T$ is a coherent integration time. When $\phi$ is a Gaussian random variable, 
the mean-square value of $C$ is defined as 
\begin{equation}
\left< C^2(T) \right> = \frac{2}{T}\int^T_0 \left( 1-\frac{\tau}{T} \right) \exp \left[ -2 \pi^2 \tau^2 \nu_0^2 I^2(\tau) \right] d\tau,
\label{eq:cohe}
\end{equation}
and the relation between the Allan variance and the true variance becomes 
\begin{equation}
\sigma^2_y(\tau) = 2\left[ I^2(\tau) - I^2(2\tau) \right],  
\label{eq:relation}
\end{equation}
where $\tau$ is the time interval, and $\nu_0$ is the arbitrary frequency before 
sampling [see Eqs. (5), (9) and (11) in \cite{rogers}]. 
When $\left< C^2(T) \right>$ is equal to 1, there is no loss of coherence. 
From the results of the phase stability measurement (Table \ref{tab:allan}) 
and the relation between the true variance and the Allan variance (Appendix A), 
the rms value of the coherence function, Eq.(\ref{eq:cohe}), can be represented as 
\begin{eqnarray}
\left< C^2(T) \right> = \frac{2}{T} \int^T_0 \left( 1-\frac{\tau}{T} \right) 
\exp \biggl[ -2 \pi^2 \tau^2 \nu_0^2  \biggl\{ \frac{2}{3}A_0^2\tau^{-2} \nonumber \\
 +B_0^2 \tau^{-1} + \frac{C_0^2}{2\log 2} \log \left( \frac{T}{\tau} \right) 
\biggl\} \biggl] d\tau
\label{eq:coherence}
\end{eqnarray}
Thus, the analysis results from the measurement of the Allan variance are of 
great help to estimate the coherence loss and the coherent integration time for 
evaluation of the ADC performance. 
Finally, the values of coherence function $\left< C^2(T) \right>$ of the developed 
ADCs are more than 0.9998 for $T=1$ sec, 0.99998 for $T=10$ sec, 0.99999 for $T=10^2$ 
sec and $T=10^3$ sec, which are derived from Eq.(\ref{eq:coherence}) by using the 
measured coefficients listed in Table \ref{tab:allan}. 

The derived Allan standard deviation shows the time fluctuation of a difference 
between two ADCs, but is not jitter time of one ADC. 
However, the Allan standard deviation can be represented with the root of the 
product of one ADC jitter time and another one. 
If two ADCs have the identical performance, the jitter time $\sigma_t$ is given by: 
\begin{equation}
\sigma_t = \frac{\tau}{\sqrt{2}} \sigma_y. 
\end{equation}
Since the white phase noise and the flicker phase noise are dominated in 1 
second as shown in Fig.~\ref{fig:09}, the jitter time is less than 0.05 psec 
from the Allan standard deviation at an averaging time of 1 second.

\subsection{Passband Response Measurement}

The sensitive and accurate astronomical observations need an analog IF system 
that has flat spectrum from the receiver to the ADCs. If there are a slope and 
a ripple, they have adverse effect on the observation. 
Several different frequency responses are caused throughout the entire analog 
processing and in ADC, and then final slope and ripple in the spectrum after 
correlation are caused by mixing these heterogeneous responses. 
For example, the slope and ripple affect the profiles of emission lines from 
astronomical objects. 

\begin{figure}
 \begin{center}
  \FigureFile(80mm,80mm){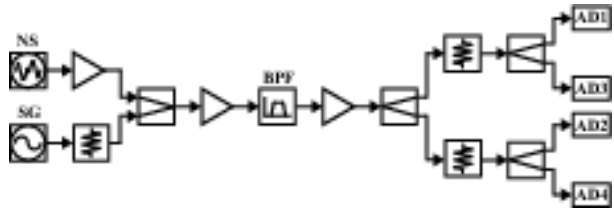}
 \end{center}
 \caption{Block diagram of the instrument setup for measurements of passband 
 responses. SG is a signal generator(Agilent Technologies E8241A).}
 \label{fig:10}
\end{figure}

The instrument setup for measuring passband responses of the ADCs is shown in 
Fig.~\ref{fig:10}. 
Input signals are made with a noise source (Agilent Technologies 346B) and a 
high-stability signal generator (Agilent E8241A) that outputs a CW signal of 
$\pm$ 1.0 dBm with accuracy. 
Power of a CW signal is as much as that of the noise source at the frequency 
channel of the correlator output. 
Frequency of the CW signal is continuously changed within the frequency range 
of the correlator outputs from 4096 MHz to 8192 MHz. 
Then, the measured spectrum can be considered as 
\begin{equation}
B_{\rm ON}(\nu) -B_{\rm OFF}(\nu) 
= B_{\rm ADB}(\nu) \times B_{\rm A}(\nu) \times B_{\rm CW}(\nu)
\label{eq:passband}   
\end{equation}
where $B_{\rm ON}(\nu)$ and $B_{\rm OFF}(\nu)$ are correlator outputs, and ON/OFF 
indicates whether the CW signal is transmitted or not. 
$B_{\rm ADB}(\nu)$, $B_{\rm A}(\nu)$, and $B_{\rm CW}(\nu)$ are passband responses of 
the ADC boards, the analog IF system, and the signal generator, respectively.  

\begin{figure}
 \begin{center}
  \FigureFile(80mm,80mm){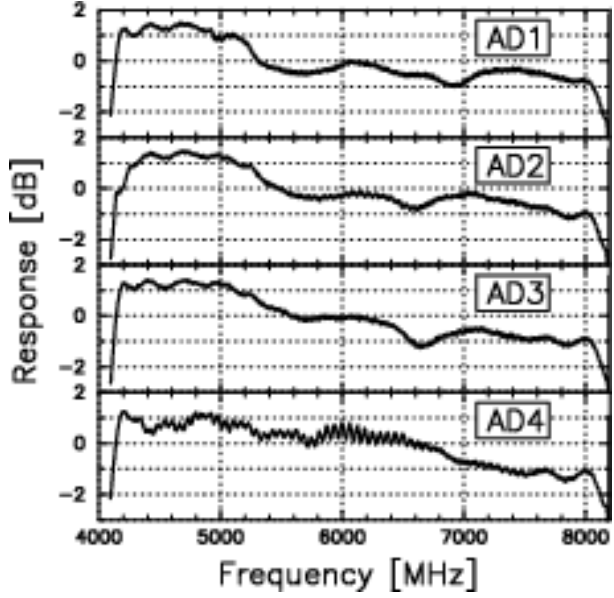}
 \end{center}
 \caption{Passband responses of the ADCs with the frequency resolution of 1 MHz. 
 The passband responses are normalized by 
 $\int^{\nu_2}_{\nu_1} B_{\rm ADB}(\nu) d\nu = 1$, 
 where $\nu_1$ and $\nu_2$ are $4097$ MHz and $8192$ MHz, respectively. }
 \label{fig:11}
\end{figure}

The passband response of the analog processing, which is obtained from the 
transmission processing of the signal between the signal generator and the ADC 
boards, can be derived with a network analyzer (Agilent E8362B) and its frequency 
range is from 3000 MHz to 9000 MHz with a frequency resolution of 1 MHz. 
With respect to $B_{\rm A}(\nu)$, since the relative accuracy is less than $0.1$ 
dB at each frequency, the synthesizer is assumed to have a flat spectrum. 
These passband responses are shown in Fig.~\ref{fig:11}. 
The measurement results are summarized in Table \ref{tab:passband}. 
In this paper, the values of ripple (p-p) is defined by subtracting the slopes 
from the passband responses. There are slopes of less than 0.73dB / GHz, which 
is similar to the specification of analog amplifiers (Quinstar Technology 
QLJ-04082530-JO) in the ADC boards. 
The peak-to-peak values of the ripple are less than 1.8 dB. However, the 
spectrum of AD4 seems to have another ripple that is less than $\pm 0.2$ dB. 
Since this ripple becomes small when an attenuator is inserted 
between the input port of the ADC boards and 
the cable connector end in the measurement system, 
it is assumed that the root cause of 
the ripple is an impedance mismatch between the ADC boards 
and this measurement system. 

Note that these results include the passband responses of the entire ADC boards 
that are composed of the amplifiers, the attenuators, semi-rigid cables, 
and DEC chips. 

\subsection{Aperture Time Measurement}

In the ADCs, the input signals are sampled during the aperture time and quantized 
for conversion to the digital signal. 
The 1-bit digitalized CW signal is expressed as 
\begin{equation}
f(t) = \sum_n \delta (t-n\frac{1}{\nu_0}) \ast g(t),  
\label{eq:f(t)}
\end{equation}
where $\nu_0$ is the CW frequency and $g(t)$ is a response of the ADC in the 
time domain. 
The Fourier transform of Eq.(\ref{eq:f(t)}) is 
\begin{equation}
F(\nu) = \nu_0 \sum_m \delta(\nu - m \nu_0) \cdot G(\nu) 
= \nu_0 \sum_m G(m\nu_0). 
\end{equation}
By measuring the transfer function, $\tilde{G}(\nu)$, a response of the ADC can 
be obtained in the time domain. 
In addition, this response provides the aperture time of the 1-bit ADC. 

\begin{figure}
 \begin{center}
  \FigureFile(80mm,80mm){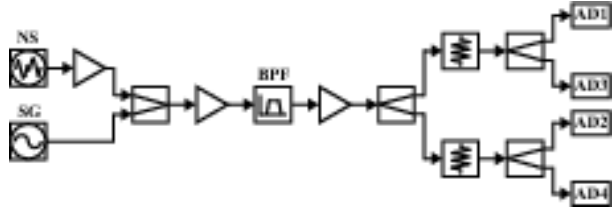}
 \end{center}
 \caption{Block diagram of the instrument setup for the measurements of the 
 aperture  times.}
 \label{fig:12}
\end{figure}

The instrument setup for measuring the aperture time of the ADCs is shown in 
Fig.~\ref{fig:12}.
Input signals are generated with the noise source and the high-stability signal 
generator which are identical with those used for the measurement of passband 
response. 
Power of the CW signal is greater than $+$40 dB, and is as much as that of the 
noise source with the 1 MHz resolution of the correlator outputs. 
The frequencies of the CW signal are set to 5.0 GHz, 6.0 GHz, and 7.0 GHz. 

\begin{figure}
 \begin{center}
  \FigureFile(160mm,160mm){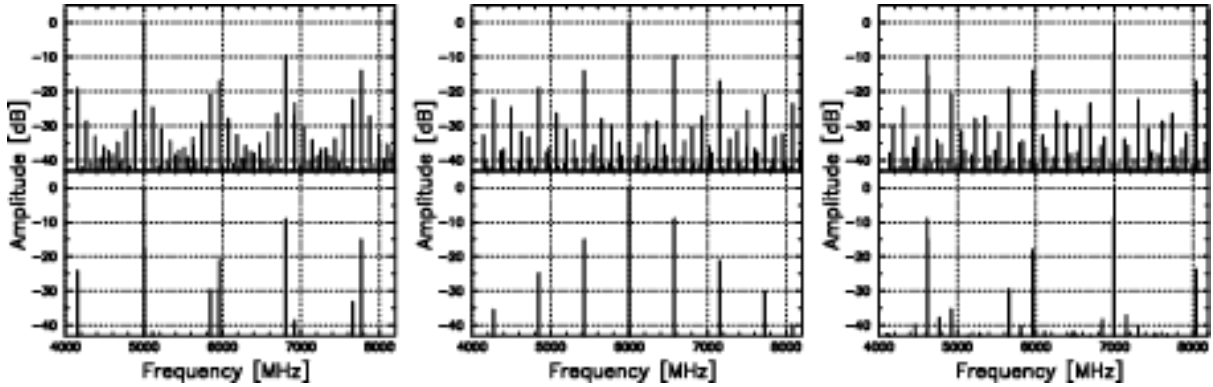}
 \end{center}
 \caption{Power spectra of CW signals, 5.0 GHz, 6.0 GHz, and 7.0 GHz with a 
 sampling frequency of 8192 MHz. 
 Amplitude is normalized by values of the amplitude of each CW signal. 
 The upper figures and the lower figures indicate the ideal spectra and the 
 measured spectra, respectively. }
 \label{fig:13}
\end{figure}

\begin{figure}
 \begin{center}
  \FigureFile(80mm,80mm){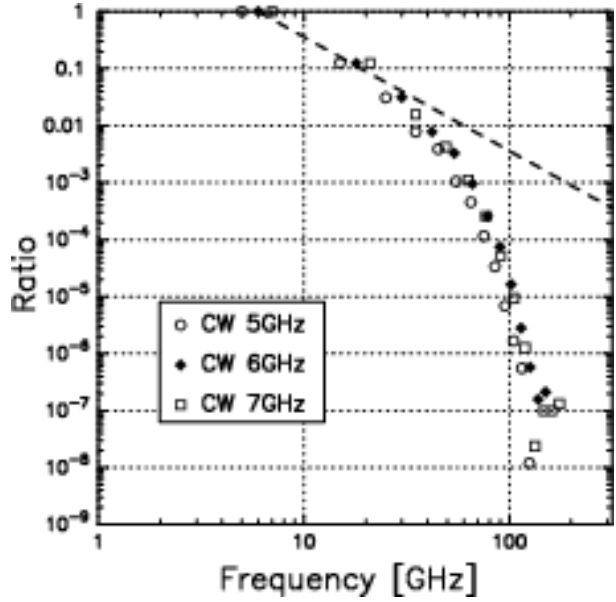}
 \end{center}
 \caption{Transfer functions for the CW frequencies of the AD1. Ratios are 
 normalized by a spectral-line power at the first-ordered frequency (5.0 GHz, 
 6.0 GHz, and 7.0 GHz).The dashed line indicates an ideal transfer function for 
 the CW frequency of 6.0 GHz. }
 \label{fig:14}
\end{figure}

Figure.~\ref{fig:13} shows a comparison between the ideal and measured power 
spectra of the digitized CW signal ($\nu_0 = 5.0, \ 6.0, \ {\rm and} \ 7.0$ GHz) 
in a sampling frequency of 8192 MHz.  
Except a CW input frequency and aliasing 
effects, the real frequencies of these spectrum line before being folded in 
sampling become $\nu = 2(n+1)\nu_0$ ($n=$1, 2, 3, ...). The transfer functions 
(see Fig.~\ref{fig:14}), $\tilde{G}(\nu)$, are derived from the power spectra. 
The result shows that the measured values of the real transfer functions become 
smaller than the ideal transfer functions when the frequency ($\nu$) is more 
than $\sim 40$ GHz.

\begin{figure}
 \begin{center}
  \FigureFile(160mm,160mm){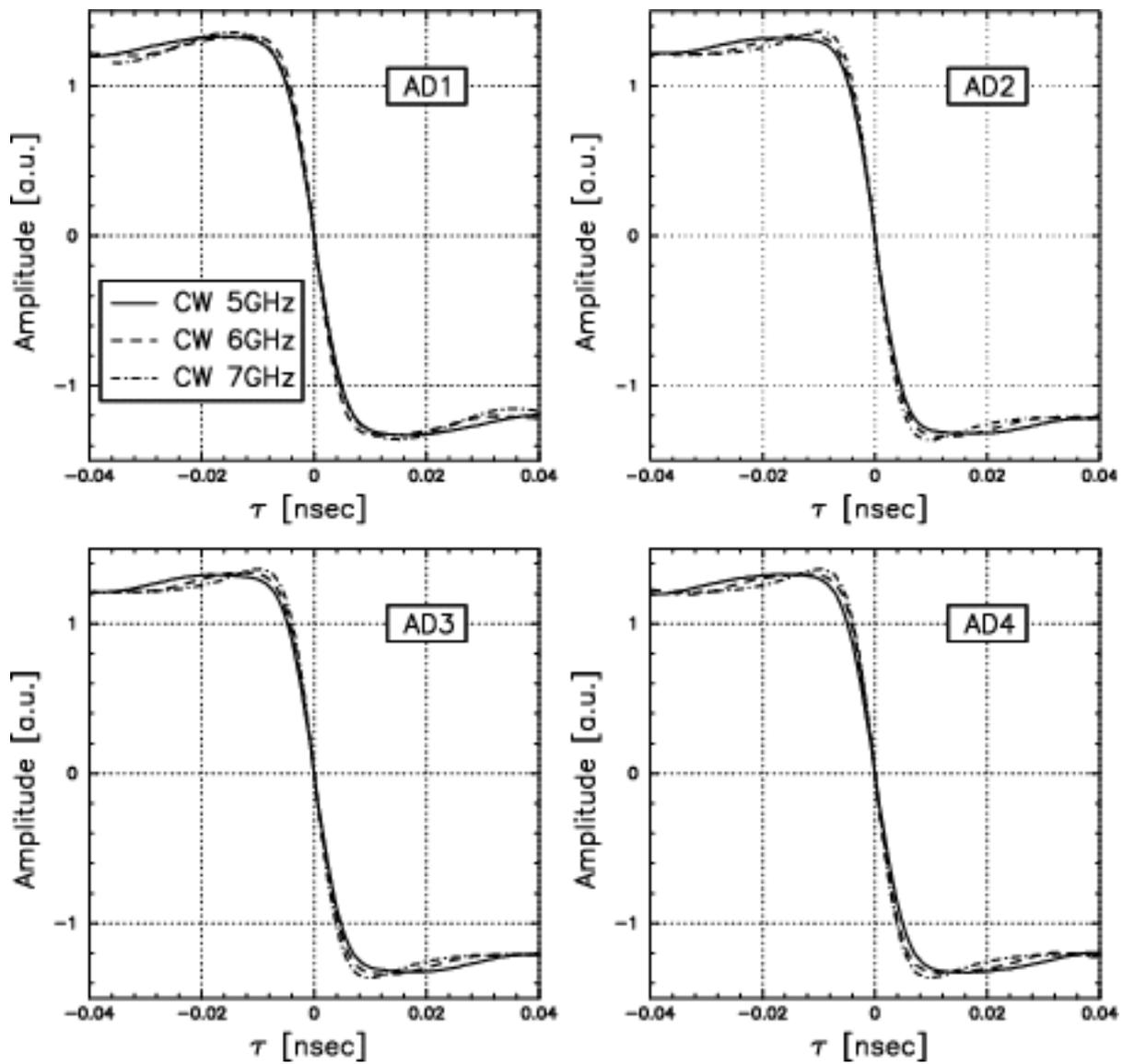}
 \end{center}
 \caption{Aperture time of the ADCs. 
 The aperture time of the ADC is less than 20 psec.}
 \label{fig:15}
\end{figure}

The frequency responses of the ADCs in the time domain are derived from the 
Fourier transform of the transfer functions below $\nu \sim 200$ GHz, and are 
shown in Fig.~\ref{fig:15}. 
Thus, the aperture time of the ADCs is less than 20 psec.

\section{Conclusion}
This paper has described the details of the developed 8-Gsps 1-bit ADCs and 
presented the measurement methods and the performances of the ADCs. Summary of 
this paper is as follows.
\begin{itemize}
\item
The ADCs were developed with the commercially 
available DEC and DMUX at a bit rate of 10 Gbps. 
Input signals with power of $-20$ dBm are quantized 
with a sampling frequency of 8192 MHz or 4096 MHz. 
\item
The Allan standard deviation of the ADC boards indicates that their performance 
is stable in the timescale of $\sim10^4$ sec. 
The jitter time in one second is estimated to be less than 0.05 psec.   
\item
There are slopes of $\leq0.73$ dB/GHz in the passband responses of 
the developed ADC boards. 
The peak-to-peak values of the ripple are less than $1.8$ dB. 
\item
The aperture time of the developed ADCs was derived from power spectra when a 
strong CW signal is input into the developed ADC boards. 
The aperture time of the ADCs is estimated to be less than 20 psec. 
\end{itemize}
The performance of the developed ADCs is summarized in Table \ref{tab:performance}. 

\section*{Acknowledgements}
We would like to acknowledge H.~Kiuchi for the useful comments on the 
calculations of the Allan variance. 
We also acknowledge S.~Asayamam, T.~Fujii, M.~Inada, and H.~Iwashita. 
We wish to thank Niizeki, Y., Fujii, Y., Takei, K., Saito, K., Ozeki, K., Kawakami, K.,  
and Onuki, H. for helpful advices about the measurements. T.~O.~ was financially 
supported by the Japan Society for the Promotion of Science (JSPS) for Young 
Scientists. 
This research was partly supported by the Ministry of Education, Culture, 
Sports, Science, and Technology, Grant-in-Aid for Young Scientists (B), 
No.~17740114, 2005 and by Grant-in-Aid for Scientific Research, No.~15037212.


\appendix
\section{The relation between the Allan variance and the true 
variance, and the coherence loss.}

It is well known that the white-phase noise has an Allan variance of 
\begin{equation}
\sigma^2_y(\tau) = A_0^2 \tau^{-2}, 
\end{equation}
and from Eq.(\ref{eq:relation}) the true variance of 
\begin{equation}
I^2(\tau) = \frac{2}{3} A_0^2 \tau^{-2}, 
\label{eq:wp}
\end{equation}
where $A_0$ is the Allan variance at an averaging time of 1 second. 
And by substituting Eq.(\ref{eq:wp}) into Eq.(\ref{eq:cohe}), 
the coherence loss in the case of white-phase noise is written as
\begin{equation}
\left< C^2(T) \right> = \exp \left( \frac{-4 \pi^2 \nu_0^2 A_0^2}{3} \right). 
\label{eq:wp-loss}
\end{equation}

Also, it is well understood that the white-frequency noise and the 
random-walk-of-phase noise have an Allan variance of 
\begin{equation}
\sigma^2_y(\tau) = B_0^2 \tau^{-1}, 
\end{equation}
and from Eq.(\ref{eq:relation}) the true variance of 
\begin{equation}
I^2(\tau) = B_0^2 \tau^{-1}, 
\label{eq:rp}
\end{equation}
where $B_0$ is the Allan variance at an averaging time of 1 second. 
And by substituting Eq.(\ref{eq:rp}) into Eq.(\ref{eq:cohe}), 
the coherence loss in the case of white-frequency noise and the 
random-walk-of-phase noise is written as
\begin{equation}
\left< C^2(T) \right> = \frac{2 ( e^{-a T} +a T -1)}{a^2 T^2}, 
\label{eq:rp-loss}
\end{equation}
where $a=2 \pi^2 \nu_0^2 B_0^2$.

The flicker-frequency noise has a constant Allan variance of  
\begin{equation}
\sigma^2_y(\tau) = C_0^2, 
\end{equation}
but the relation between Allan variance and the ture variance 
has been directly solved and expressed yet \cite{thompson,rogers}. 
However, by defining the finite limited time in Eq.(\ref{eq:relation}), 
the true variance can be derived as 
\begin{equation}
I^2(\tau) = \frac{C_0^2}{2 \log 2}\log\left( \frac{\tau_{max}}{\tau} \right), 
\end{equation}
where $\tau_{max}$ is a constant of an arbitrary value to be fixed by 
introducing some constraints. 
For instance, in the atmospheric phase instability, the Allan variance 
profile changes from the flicker-frequency noise to the white-phase 
noise and the flicker-phase noise in the time range of 10 to 100 seconds. 
Also, the Allan variance profile changes from the flicker-frequency noise 
to the random-walk-of-frequency noise in the atomic frequency standard. 
Thus, it can be defined that $\tau_{max}$ is the turnover time of Allan 
variance profile. 
If the coherent time of $T$ is more than $\tau_{max}$, the ture variance 
becomes 
\begin{equation}
I^2(\tau) = 0, 
\end{equation}
while if $T$ is less than $\tau_{max}$,  
\begin{equation}
I^2(\tau) = \frac{C_0^2}{2 \log 2}\log\left( \frac{T}{\tau} \right). 
\label{eq:fp}
\end{equation}
And by substituting Eq.(\ref{eq:fp}) into Eq.(\ref{eq:cohe}), 
the coherence loss in the case of white-phase noise and flicker-phase noise 
is written as
\begin{equation}
\left< C^2(T) \right> = \frac{2}{T} \int^T_0 \left( 1-\frac{\tau}{T} \right) 
\exp \biggl[ -\frac{\pi^2 \tau^2 \nu_0^2 C^2_0}{\log 2} 
\log \left( \frac{T}{\tau} \right) \biggl]  d\tau, 
\label{eq:fp-loss}
\end{equation}

The random-walk-of-frequency noise has an Allan variance of 
\begin{equation}
\sigma^2_y(\tau) = D_0^2 \tau, 
\end{equation}
where $D$ is the Allan variance at an averaging time of 1 second. 
By using the same derivation as that of the flicker-frequency noise, 
it is also defined $\tau_{end}$ is a finite integration time.
When the coherent time of $T$ is more than $\tau_{end}$, 
the ture variance becomes 
\begin{equation}
I^2(\tau) = 0, 
\end{equation}
while when $T$ is less than $\tau_{end}$,  
\begin{equation}
I^2(\tau) = \frac{D_0^2}{2} (T - \tau). 
\label{eq:rf}
\end{equation}
And by substituting Eq.(\ref{eq:rf}) into Eq.(\ref{eq:cohe}). 
the coherence loss in the case of random-walk-of-frequency noies is 
written as
\begin{equation}
\left< C^2(T) \right> =  \frac{2}{T} \int^T_0 \left( 1-\frac{\tau}{T} \right) 
\exp \biggl[ -\pi^2 \nu_0^2 D^2_0 \tau^2 (T-\tau) \biggl] d\tau
\label{eq:rf-loss}
\end{equation}

\newpage


\newpage

\begin{table}
 \begin{center} 
  \caption{The composition of 1:2(4) DMUX and the observation data according to 
 the sampling rates.}
  \begin{tabular}{ccc}
   \hline\hline
Sampling rate & Decimation & Description \\ 
\hline 
8Gsps & 1:4 & Either of IF1 or IF2 data \\
4Gsps & 1:2 & 64 bits each for IF1 and IF2 \\
\hline
  \end{tabular}
 \end{center}
 \label{tab:sampling}
\end{table}%
\begin{table}
 \begin{center} 
  \caption{ The Allan variances in 1 second for various types of noise. 
   Time rage indicate fitting ranges.} 
  \begin{tabular}{cccccc}
   \hline\hline
   Correlation & Freq [GHz] & $A_0$ & $B_0$ & $C_0$ & Time range [sec] \\
   \hline
   AD1$\ast$AD2 & $5.0$ & $6.3\times10^{-14}$ & $2.2\times10^{-15}$ & $6.5\times10^{-17}$ & $1.0 \le \tau \le 9.0\times10^2$ \\
                &       & $1.1\times10^{-13}$ & ---                 & ---                 & $9.0\times10^2 < \tau  \le 1.5\times10^3$ \\
                & $6.0$ & $5.4\times10^{-14}$ & $2.0\times10^{-15}$ & $2.8\times10^{-17}$ & $1.0 \le \tau \le 9.0\times10^2$ \\
                &       & $8.3\times10^{-14}$ & ---                 & ---                 & $9.0\times10^2 < \tau  \le 1.5\times10^3$ \\
                & $7.0$ & $5.0\times10^{-14}$ & $1.4\times10^{-15}$ & $5.6\times10^{-17}$ & $1.0 \le \tau \le 9.0\times10^2$ \\
                &       & $8.0\times10^{-14}$ & ---                 & ---                 & $9.0\times10^2 < \tau  \le 1.5\times10^3$ \\
   \hline 
   AD3$\ast$AD4 & $5.0$ & $5.7\times10^{-14}$ & $3.2\times10^{-15}$ & $1.5\times10^{-16}$ & $1.0 \le \tau \le 1.3\times10^3$ \\
                &       & $2.3\times10^{-13}$ & ---                 & ---                 & $1.3\times10^3 < \tau  \le 2.3\times10^3$ \\
                & $6.0$ & $5.0\times10^{-14}$ & $3.1\times10^{-15}$ & $1.4\times10^{-16}$ & $1.0 \le \tau \le 1.3\times10^3$ \\
                &       & $2.2\times10^{-13}$ & ---                 & ---                 & $1.3\times10^3 < \tau  \le 2.3\times10^3$ \\
                & $7.0$ & $6.3\times10^{-14}$ & $4.0\times10^{-15}$ & $1.8\times10^{-17}$ & $1.0 \le \tau \le 1.3\times10^3$ \\
                &       & $2.8\times10^{-13}$ & ---                 & ---                 & $1.3\times10^3 < \tau  \le 2.3\times10^3$ \\
   \hline
   \end{tabular}
 \label{tab:allan}
 \end{center}
\end{table}%
\begin{table}
 \begin{center}
  \caption{ Passband response for ADC boards.} 
  \begin{tabular}{ccc}
   \hline\hline
   ADC bord & slope [dB/GHz] & ripple(p-p) [dB]\\
   \hline
   AD1 & 0.58 & 1.6 \\
   AD2 & 0.63 & 1.4 \\
   AD3 & 0.73 & 1.8 \\
   AD4 & 0.64 & 1.5 \\
   \hline
   \end{tabular}
 \label{tab:passband}
 \end{center}
\end{table}%
\begin{table}
 \begin{center}
  \caption{ Performance of the developed ADCs.} 
  \begin{tabular}{ll}
   \hline\hline
   Sampling Frequency       & 4096 MHz or 8192 MHz\\
   Input level              & $-20$ dBm \\
   Number of bit            & 1 \\
   Jitter time in 1 second  & $<$0.05 psec\\
   Passband response        &  \\
   \ \ slope                & $\leq 0.73$ dB/GHz \\
   \ \ ripple(p-p)          & $\leq 1.8$ dB\\
   Aperture time            & $<$ 20 psec \\
   \hline
   \end{tabular}
 \label{tab:performance}
 \end{center}
\end{table}%
\end{document}